\documentclass[reprint,prb,showpacs,preprintnumbers,amsmath,amssymb,twocolumns,footinbib,citeautoscript]{revtex4-1}

\usepackage{graphicx}
\usepackage{dcolumn}
\usepackage{bm}

\begin{document}

\title{Understanding the Electronic Transport Through Single Noble Gas Atoms}

\author{L. A. Zotti,$^1$ M. B\"urkle,$^2$ Y. J.  Dappe,$^3$ F. Pauly,$^2$ and J. C. Cuevas$^{1}$}

\affiliation{$^1$Departamento de F\'{\i}sica Te\'orica de la Materia Condensada,
Universidad Aut\'onoma de Madrid, E-28049 Madrid, Spain \\
$^2$Institut f\"ur Theoretische Festk\"orperphysik and DFG Center for Functional 
Nanostructures, Karlsruhe Institute of Technology (KIT), D-76131 Karlsruhe, Germany \\
$^3$Institut de Physique et Chimie des Mat\'eriaux de Strasbourg (IPCMS), UMR CNRS 7504, 
23 rue du Loess, POB 43, F-67034 Strasbourg Cedex, France}

\date{\today}

\begin{abstract}
We present a theoretical study of the conductance of atomic junctions comprising
 single noble gas atoms (He, Ne, Ar, Kr, and Xe) coupled to gold electrodes.
The aim is to elucidate how the presence of noble gas atoms affects the electronic transport 
through metallic atomic-size contacts. Our analysis, based on density functional theory and 
including van der Waals interactions, shows that for the lightest elements (He and Ne) 
no significant current flows through the noble gas atoms and their effect is to reduce 
the conductance of the junctions by screening the interaction between the gold electrodes. 
This explains the observations reported in metallic atomic-size contacts with adsorbed He 
atoms. Conversely, the heaviest atoms (Kr and Xe) increase the conductance due to 
the additional current path provided by their valence $p$ states. 
\end{abstract}


\maketitle
Noble gases  are commonly employed in scanning probe experiments as exchange gases 
since they are expected to interact weakly with the studied systems. Furthermore, it is 
often assumed that the adsorption of NG atoms does not affect the electron tunneling 
between metallic electrodes. However, it has been shown that this is not entirely true. 
For instance, two decades ago Eigler and coworkers presented scanning tunneling
microscope (STM) images of Xe atoms on a Ni(110) surface \cite{Eigler1991} and they
nicely demonstrated that these atoms can be moved to chosen positions on the surface. 
It has also been shown that it is possible to manipulate individual Xe atoms to 
construct atomic wires and to measure their electrical resistance \cite{Yazdani1996} 
or to functionalize molecules \cite{Qiu2002}.

From the theory side, while there are numerous works analyzing the interaction between 
noble gas (NG) atoms and metal surfaces, studies exploring the transport through
metal-NG-metal junctions are rather scarce, and most of them have focused either
on understanding atomic manipulation or on  STM imaging
\cite{Yazdani1996,Lang1986,Mahanty1991,Saenz1993,Lang1994,Flores1995}. 
There are still important open problems concerning how adsorbed NG atoms modify the 
transport through metallic atomic-size junctions. A striking example is the observation
made in several break-junction and STM experiments that adsorbed 
He atoms can strongly modify the current through metallic junctions, lowering 
in particular the low-bias conductance \cite{Keijers1996,Kolesnychenko1999,Keijers2000,
Untiedt2002}. This conductance suppression is surprising since the height of 
the tunneling barrier in the presence of NG atoms has been predicted to decrease 
\cite{Mahanty1991} and, indeed, Kelvin probe experiments have shown that the work function 
of noble metal surfaces decreases upon adsorption of Ar, Kr and Xe \cite{Huckstadt2006}. 
A possible explanation, based on predictions by Lang \cite{Lang1986}, 
suggests that adsorbed He atoms can polarize metal states away from the Fermi energy, 
leading to a decrease in the metal local density of states. This explanation was based 
on calculations where the metal electrodes were described by a jellium model (with no 
atomistic details) and without taking into account van der Waals interactions. Thus, it 
is highly desirable to revisit this problem with ab initio transport methods.

To shed new light on the influence of adsorbed NG atoms in the transport through 
metallic atomic contacts, we present in this Brief Report a systematic ab initio study of the 
conductance of gold atomic junctions containing single atoms of He, Ne, 
Ar, Kr, and Xe. Our calculations, based on density functional theory (DFT), show that, 
while for He and Ne the current flows directly from one metallic electrode to the 
other, for Ar, Kr and Xe the transport occurs mainly through the valence $p$ states of 
the NG atom. In all cases, the presence of NG atoms induces a dipole moment which screens
the interaction between the leads. In the case of He, Ne, and Ar the weakening of the
metal-metal coupling (rather than a suppression of the metal density of states,
as proposed by Lang \cite{Lang1986}) leads to a reduction of the tunneling current. 
On the contrary, for Kr and Xe the additional tunneling path provided by the valence 
$p$ states overcomes the screening, leading to an enhancement of the current.

\begin{figure}[t]
\begin{center}
\includegraphics[width=8cm]{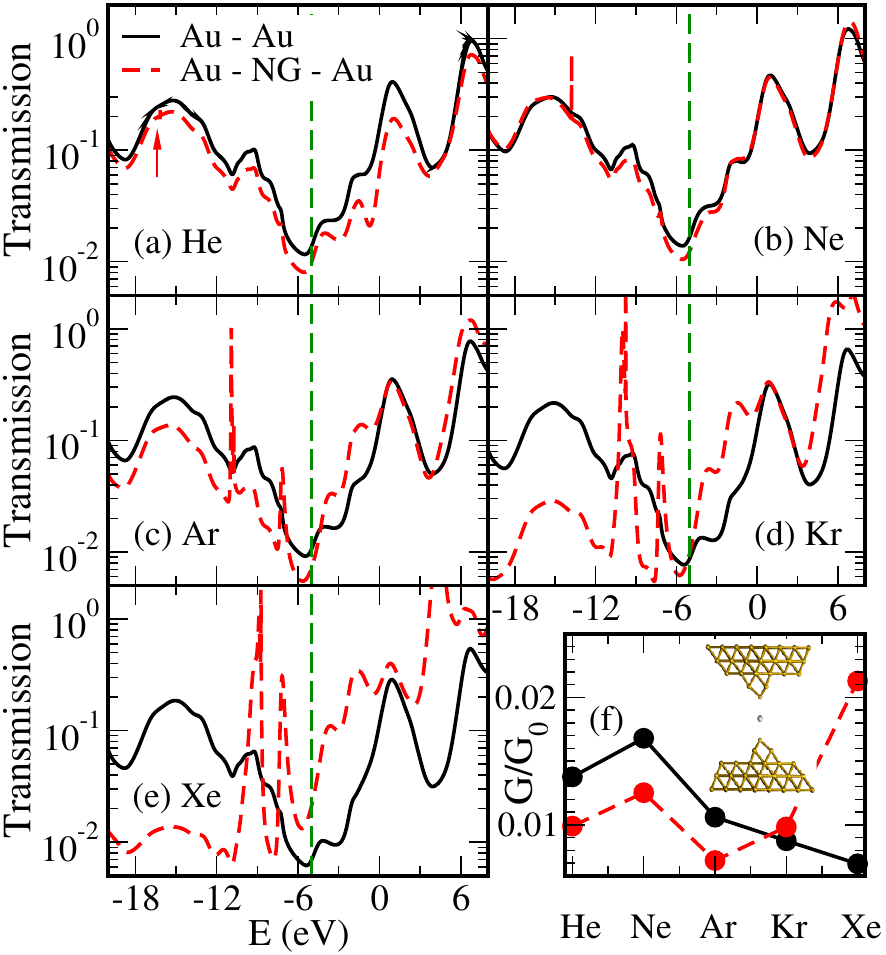}
\caption{(Color online) (a-e) Zero-bias total transmission as a function of energy for 
Au-Au (solid black) and Au-NG-Au (dashed red) junctions in the top binding
geometry, as shown in the inset of panel (f). The vertical dashed lines indicate
 the gold Fermi level (-5 eV). (f) Comparison of conductances for Au-Au and Au-NG-Au
 junctions for all studied NG atoms.}
\label{trans-NG}
\end{center}
\end{figure}

Our main goal is to analyze the electronic transport through metallic atomic-size 
contacts containing single atoms of noble gases. In particular, we have chosen
gold for the electrode material and studied the elements He, Ne, Ar, Kr, and Xe. 
For this purpose, we have carried out conductance calculations within the framework 
of DFT following the method described in Ref.~[\onlinecite{Pauly2008}],
 which is built upon the Turbomole 6.1 code \cite{Turbomole}.
 In all our calculations we have used the BP86 functional \cite{BP86}. 
The first step in our analysis is the construction of the atomic junctions. This
is done by optimizing geometries where the gold electrodes are formed by two finite
clusters of 20 atoms and a single NG atom is placed in the middle. In the 
optimization, the NG atom and the four innermost gold atoms on each side were 
relaxed, while the other gold atoms were kept frozen. For the optimized atoms, a 
def2-TZVP basis set \cite{Weigend2005} was chosen, while a def-SVP basis set 
\cite{Schafer1992} was used for the frozen gold atoms. The binding energies 
calculated in this way were found to differ by only around $5 \times 
10^{-4}$ eV from those calculated with a def2-TZVP basis set for all the atoms. 
Subsequently, the gold cluster size was extended to 116 atoms on each side in 
order to describe the metal-NG atom charge transfer and the energy level alignment 
correctly (see inset in Fig.~\ref{trans-NG}). It is important to emphasize that we 
have used the semiempirical DFT-D2 correction \cite{Grimme2006} in order to take 
the dispersive forces into account, since the binding distance between NG and noble 
metals is known to be determined by the interplay between the Pauli repulsion and the 
van der Waals interaction \cite{Pershina2008,Muller2007}. Finally, the information 
about the electronic structure of the junctions obtained with DFT is transformed into
linear conductance using Green's function techniques as described in detail 
in Ref.~[\onlinecite{Pauly2008}]. This is done in the spirit of the Landauer 
approach, where the low-temperature linear conductance is given by $G= 
G_0 T(E_{\rm F}) = G_0 \sum_i \tau_i(E_{\rm F})$, where $G_0=2e^2/h$ is
the quantum of conductance, $T(E_{\rm F})$ is the total transmission of the junction 
at the Fermi energy, $E_{\rm F}$, and $\{ \tau_i \}$ are the transmission
coefficients, i.e., the eigenvalues of the transmission matrix.

We now start our analysis of the results by comparing the linear
conductance of gold junctions containing the five NG atoms considered in this work.
We consider firstly junctions with a top binding position
(see inset in Fig.~\ref{trans-NG}(f)), since such a geometry 
has been suggested as the most favorable for most of the NG atoms on metal surfaces
\cite{Pershina2008,Petersen1996,Diehl2004,DaSilva2005,DaSilva2008}.
The Au-NG-atom distances and the corresponding binding energies are listed
in Table~\ref{table1}. In particular, the binding energies increase as we move 
to heavier elements, in agreement with the calculations of Pershina \emph{et al.}\ 
\cite{Pershina2008}. In Fig.~\ref{trans-NG}(a-e) we show the zero-bias 
transmission as a function of energy for the five Au-NG-Au junctions with top 
binding geometries (dashed lines). Notice that below $E_{\rm F}$ (marked by 
a vertical dashed line), pronounced peaks appear, which move towards the Fermi energy
as the atomic number of the NG atom increases. These peaks appear approximately
at the energies of the highest occupied states of the NG atoms in the gas phase (see
Table~\ref{table1}), which suggests that they originate from the valence $p$ states 
for Ne, Ar, Kr, and Xe, and from the $1s$ state for He. A closer look at the peaks 
for Ne, Ar, Kr, and Xe shows that they are split into two. One corresponds to the 
$p_z$ orbital ($z$ being the direction of the junction axis), which is shifted
to lower energies, and the second is due to the $p_x$ and $p_y$ states,
which remain degenerate in the junction \cite{SI}. The peak or resonance due to 
the $p_z$ orbital is clearly broader due to its stronger hybridization with the 
gold states. Moreover, the width of this resonance increases from Ne to Xe
simply because it is determined by the local density of states (LDOS) of the 
 gold tip atoms at the energy of the valence states of the NG atoms. For Ne 
and Ar, that energy lies outside the $5d$ band of gold, while for Kr and Xe, it
is well inside this band. Regarding the conductance, displayed in 
Fig.~\ref{trans-NG}(f), it varies in a non-monotonic manner from $10^{-2}G_0$
for He to approximately $2.1 \times 10^{-2}G_0$ for Xe. Notice that 
 the computed conductance for Xe is lower by one order of magnitude than that
measured in Ref.~[\onlinecite{Yazdani1996}]. We attribute this to the different
 electrode material (Au rather than Ni).

\begin{table}[t]
\begin{tabular}{|c|c|c|c|c|c|c|}
\hline 
     &              &              & Au-NG     & Dipole  & Charges on & Binding \\
Atom & $\epsilon_h$ & $\epsilon_l$ & distance  & moment  &   NG atom  & energy  \\
     &  (eV)        & (eV)         & (\AA)     & (debye) &   ($|e|$)  & (eV)    \\
\hline
He & -15.79 & 16.69 & 3.41 & 1.26 & 0.019 & -0.01 \\
Ne & -13.29 & 14.99 & 3.28 & 1.38 & 0.018 & -0.05 \\
Ar & -10.30 & 10.20 & 3.55 & 1.66 & 0.033 & -0.09 \\
Kr & -9.36  & 6.81  & 3.63 & 1.76 & 0.028 & -0.13 \\
Xe & -8.32  & 4.69  & 3.73 & 2.11 & 0.028 & -0.18 \\
\hline
\end{tabular}
\caption{The second and third columns show the energy of highest occupied  
($\epsilon_h$) and lowest unoccupied ($\epsilon_l$) states for the NG atoms in 
the gas phase. The others contain calculated quantities for the top-binding 
geometries with a single gold cluster of 116 atoms: binding distance, dipole 
moment, charge on the NG atom, and binding energy. The value of the dipole 
in the bare Au cluster is 1.10 debye. Concerning the corresponding quantities
in the junction, the binding distances do not change, while the charges on the 
NG atom and the binding energies are approximately doubled.}
\label{table1}
\end{table}

To understand the mechanism governing the conduction through the NG atoms and how they 
modify the transport through the gold junctions, we also present in Fig.~\ref{trans-NG}(a-e)
the transmission curves for Au-Au junctions (with no NG atoms) and in panel (f) the 
corresponding conductance. In these calculations we have kept the gold electrodes at the 
same distance as in the corresponding Au-NG-Au junctions. The first thing to notice is 
that for the lightest elements the conductance is lowered when the NG atoms are in the 
junctions.  As explained in the introduction, such a reduction of the conductance 
caused by the adsorption of He has been observed in several low-temperature break-junction 
and STM experiments \cite{Keijers1996,Keijers2000,Untiedt2002}. 
A similar behaviour has been observed for closed shell molecules such as H$_2$ \cite{Weiss2010}.
For He and Ne, we find that the conductance is decreased by about 30\% when these 
atoms are present. On the contrary, the presence of Kr and Xe in the junctions gives
rise to an increase of the conductance, which is larger than a factor of 2 for Xe.
What is the reason for this different behavior? A first hint is obtained by 
comparing the transmission curves of the junctions with and without NG atoms. As one can 
see in Fig.~\ref{trans-NG}(a-b), for He and Ne the transmission curves for the Au-NG-Au 
junctions follow very closely the energy dependence of the Au-Au junctions. This fact 
suggests that for He and Ne, the main contribution to the current in the
Au-NG-Au junctions comes from direct tunneling from gold to gold, and the only effect 
of these NG atoms is to reduce the transmission of the gold-to-gold current path. On the 
other hand, for Ar, Kr and Xe, the transmission curves differ markedly from those of the
Au-Au junctions, especially close to the Fermi energy. This suggests that the valence 
$p$ states of these atoms are contributing significantly to the transport through these 
junctions.

\begin{figure}[t]
\begin{center}
\includegraphics[width=8.5cm]{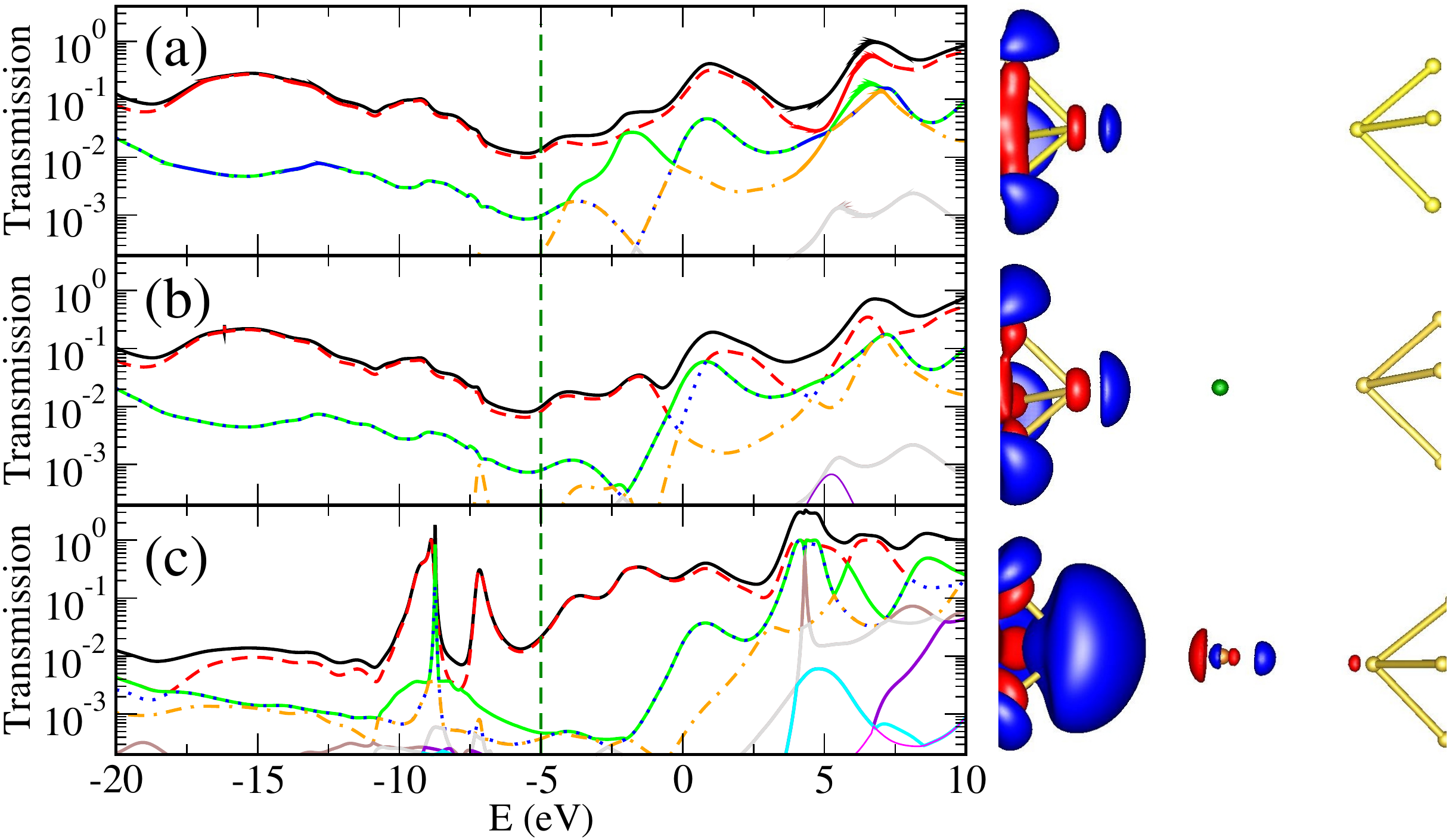}
\caption{(Color online) The left panels show the total transmission (solid lines)
and the largest transmission coefficients (dashed lines) as a function of energy
for Au-Au (a), Au-He-Au (b) and Au-Xe-Au (c) junctions. The transmission coefficients
have been classified according to the magnitude of the eigenvalue. The right panels show the
corresponding right moving wavefunctions of the dominant eigenchannel at the Fermi
energy for the same isosurface value.}
\label{fig:channel}
\end{center}
\end{figure}

The fact that He and Ne do not contribute to the electronic transport is reasonable, 
as their highest occupied states lie very far away from the Fermi level and they are 
weakly coupled to the electrodes. The conductance suppression is, however, 
surprising, as it is, a priori, not obvious how a closed-shell atom can conduct less 
than the vacuum. We explain this suppression as follows. In our analysis of the 
interaction between a NG atom and a single gold cluster we have found that there is
a tiny charge transfer (a fraction of an electron leaves the NG atoms), which induces 
a dipole moment pointing away from the metal \cite{note1}. The values of the charge 
transfers and the dipole moments are reported in Table~\ref{table1}. The existence
of this dipole moment has been predicted by numerous authors in the context of
NG atoms adsorbed on metal surfaces \cite{Muller1990,Saenz1993,Widom2000,
Pershina2008,DaSilva2008,Trioni2009}. Also in that context, it has been argued that
as a consequence of the induced dipole the charge is pushed from the interstitial vacuum
region toward the Au bulk (so-called pillow or cushion effect). In our case, we 
did not find a notable charge depletion on the gold tips. We also did not find any 
significant change in the LDOS of the gold tips at the Fermi level, as  was proposed 
by Lang \cite{Lang1986} (see Ref.~[\onlinecite{SI}]) or more recently by Weiss et al.
\cite{Weiss2010} for the H$_2$ case. Therefore, we suggest that, at least for the He
and Ne case,  what 
causes the current suppression is the induced polarization which decreases the coupling 
between the two leads, reducing the transmission of the direct gold-to-gold path. The 
discrepancy between our results and those of Ref.~[\onlinecite{Lang1986}] could be due 
to the limitations of the jellium model used in that work.

Convincing evidence of the validity of our arguments above is provided by the analysis 
of the conduction channels. In the left panels of Fig.~\ref{fig:channel} we show both 
the total transmission and the individual transmission coefficients $\{ \tau_i \}$ as 
a function of energy for the Au-Au, Au-He-Au and Au-Xe-Au junctions. In the Au-Au 
junction the interelectrode distance is the same as in the Au-He-Au junction. In all 
cases, the transmission at the Fermi energy is dominated by a single channel. More 
importantly, the transmission coefficients are very similar for the Au-Au and the 
Au-He-Au junctions, suggesting a common transport mechanism, whereas for the Au-Xe-Au 
junction new features are visible around $E_{\rm F}$, originating from valence $p$ 
states of the Xe atom. The nature of the dominant channel at $E_{\rm F}$ can be 
established by looking at the corresponding wavefunctions of this eigenchannel. Such 
(right moving) wavefunctions for these three junctions are shown in the right panels of 
Fig.~\ref{fig:channel}. They have been calculated using the method described in 
Ref.~[\onlinecite{Paulsson2007}]. Notice that there is no weight present in the gap 
region either in the Au-Au junctions, as expected, or in the Au-He-Au. On the other hand,
in the Au-Xe-Au junction the channel in the central region has the symmetry of the Xe 
$p_z$ orbital, demonstrating that this orbital plays the main role in the conduction 
through this junction, as hinted in Ref.~[\onlinecite{Pizzagalli1997}].
 This is at variance with the calculations of Ref.~[\onlinecite{Lang1994}], 
where the transport through Xe was found to occur through the tail of the $6s$ orbital.

\begin{figure}
\begin{center}
\includegraphics[width=6cm]{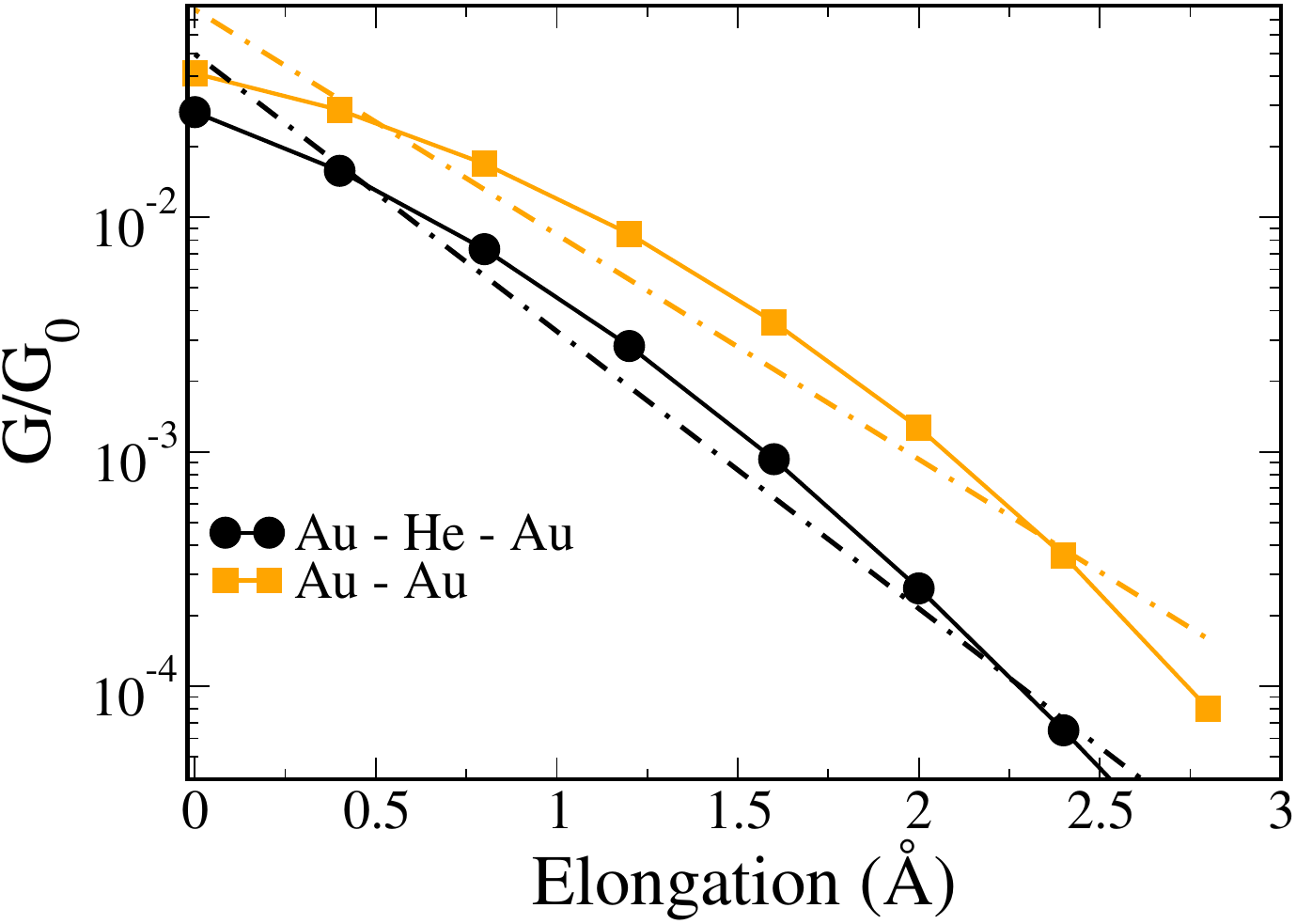}
\caption{(Color online) Conductance vs.\ elongation for Au-Au and Au-He-Au
junctions upon stretching (solid lines) and fitting curves to an exponential 
function (dashed-dotted lines), see text.}
\label{stretching}
\end{center}
\end{figure}

We have checked that the qualitative behavior of the different NG atoms is also
exhibited in other binding geometries. An example of Au-He-Au and Au-Xe-Au 
junctions with hollow binding geometries is shown in Ref.~[\onlinecite{SI}]. On the other 
hand, one may wonder whether the special behavior of the He is sensitive 
to the interelectrode distance. This is an important question since experimentally 
it is not easy to determine absolute distances, and the influence of the He atoms 
has been deduced from the conductance decay when the junctions are stretched toward 
the tunnel regime. To answer this question, we have simulated the stretching of a
Au-He-Au junction starting from the equilibrium geometry of Fig.~\ref{trans-NG}. 
For these calculations we have modeled the narrowest part of the electrodes with 
clusters of 20 gold atoms, and have checked that the results are consistent with 
those obtained with the larger clusters. In Fig.~\ref{stretching}
we show the evolution of the conductance of the Au-He-Au junction upon separating 
the gold electrodes symmetrically from the He atom in the equilibrium position.
 For comparison we also include the results for the corresponding Au-Au junction.
 Fig.~\ref{stretching} shows that 
the presence of the He atom suppresses the conductance also at larger distances. 
The conductance decay for both junctions can be roughly fitted with an
exponential function $G=A e^{-\beta d}$, where $\beta$ is the attenuation factor 
and $d$ corresponds to the elongation, as measured with respect to the equilibrium 
position. The attenuation factor $\beta$ is equal to 2.72 \AA$^{-1}$ for the 
junction with the He atom and 2.21 \AA$^{-1}$ for the pure gold junction.
The faster decay with the He atom is consistent with the experiments
\cite{Keijers1996,Keijers2000,Untiedt2002}, and it shows that the conductance 
suppression can be more dramatic at large distances (up to 82\%).

In summary, we have presented a theoretical analysis of the influence of NG 
atoms on the conductance of gold atomic contacts. We have shown that He and 
Ne atoms do not conduct and their effect is to reduce the conductance of the 
metallic junctions due to polarization effects. As we consider the tunneling 
current through heavier noble gas atoms, the polarization effects increase but 
they are compensated for by an increasing contribution of the valence $p$ 
states of the NG atoms, leading to an enhancement of the conductance. Our study 
shows that the presence of absorbed NG atoms can modify significantly the 
conduction through atomic-scale junctions, which has important implications
for a great variety of experiments in the field of molecular electronics.\\

We thank E. Leary, N. Agra\"{\i}t and T. Frederiksen for useful discussions.
L.A.Z. and J.C.C. were funded by the EU through BIMORE (MRTN-CT-2006-035859) 
and by the Comunidad de Madrid through the program NANOBIOMAGNET S2009/MAT1726.
M.B acknowledges funding through the Center for Functional Nanostructures
 and the DFG priority program 1243, and F.P. 
through the Young Investigator Group.



\begin{thebibliography}{10}
\bibitem{note1}
The total dipole moment vanishes in the Au-NG-Au junctions because of the 
symmetry of the contacts.
\end{thebibliography}
\end{document}